# Near Infrared Diffuse Interstellar Bands Characterized by Fullerene and Graphene Molecules

Norio Ota

Graduate school of Pure and Applied Sciences, University of Tsukuba, *Japan*

https://www.researchgate.net/profile/Norio-Ota-Or-Ohta

Astronomical near-infrared Diffuse-Interstellar-Bands (DIBs) were characterized by pure carbon Fullerene and Graphene molecules comparing with laboratory experiment and with Time-Dependent Density-Functional-Theory (TD-DFT) analysis. It is well known that two large DIBs of Fullerene cation (C60)+ at 9577A and 9632A coincide well with laboratory experiments. Those are thought to be split bands by the Jahn-Teller molecular deformation. In our TD-DFT calculation, those are reproduced by degenerated bands at 9549A and 9552A before deformation. Cation enriching experiment by Strelnikov et al. suggested longer wavelength two bands of DIB10542 and DIB10610 (observed by Hamano et al.), which may split from calculated 10410A and 10411A. Also, we noticed shorter wavelength experimental band around 8550A, which may relate to calculated 8677A and 8686A. We challenged such analysis on Graphene molecules as like (C54) (C53) (C52) and (C51), which are carbon hexagon and pentagon combined molecules. Calculation could reproduce many near-infrared bands. Calculated bands of (C54) suggest that one DIB among (DIB9577, DIB9632, or DIB9673) may correspond to one of (DIB10361, DIB10394, or DIB10439). Calculated bands of (C51) suggest that one of (DIB9686, DIB9987, or DIB10006) may relate to one of (DIB10262 or DIB10288). Combining astronomical observation, laboratory experiment, and quantum chemical analysis, we could suggest carrier candidates of DIBs.

## 1. Introduction

This paper predicts that near-infrared Diffuse Interstellar Bands (DIBs) [1] could be characterized by pure carbon fullerene and graphene molecules. We like to compare astronomically observed near-infrared DIBs with laboratory experiments and with molecular quantum calculation based on the Time Dependent Density Functional Theory (TD-DFT). In our previous papers [2, 3], we pointed out that astronomically observed Infrared spectrum (IR) could be reproduced well by DFT. Molecules were polycyclic pure carbon molecules as like soccer ball like fullerene-($C_{60}$), and planer configuration graphene molecules of ($C_{53}$), ($C_{52}$), ($C_{51}$), ($C_{23}$), ($C_{22}$) and ($C_{21}$) having few carbon pentagon sites among hexagon networks. In this study, we like to apply above carbon molecules to reproduce DIBs using TD-DFT [4, 5].

DIBs are a large set of absorption features at a wide range of visible to infrared wavelengths to associate with pure carbon and /or hydrocarbon molecules as reviewed by T. R. Geballe [1]. The discovery of the first DIBs was made by Mary Lea Heger [6], almost 100 years ago. Until now, there observed over 500 DIBs [7]-[12] However, any specific molecule had not been identified until 2015. After long years efforts [13]-[15], five near-infrared DIBs matched to the laboratory experimental spectrum by singly cationic fullerene ($C_{60}$)+ in 2015-2016 by E. Cambell et al. [16],[17]. However, other many DIBs were not identified yet. This is the great mystery for the astrochemistry and molecular science. Recently, systematic near infrared DIBs were reported by Hamano et al. [18]. It was believed that DIB may come from molecular orbital photoexcitation absorbed bands. In this study, we like to compare observed DIBs with laboratory photo-absorbed experiments and with TD-DFT analysis. We like to apply the specified fullerene and graphene molecules already studied by molecular vibrational IR. Purpose of this study is to indicate the specific pure carbon molecules to satisfy both astronomically observed IR and DIBs.

## 2. Calculation Methods

In calculation, we used DFT [19, 20] and TD-DFT [4, 5] with the unrestricted PBEPBE functional [21, 22]. We utilized the Gaussian09 software package[23] employing an atomic orbital 6-311G basis[24]. Unrestricted DFT calculation was done to have the spin dependent atomic structure. The required convergence of the root-mean-square density matrix was $10^{-8}$.

For obtaining the molecular orbital excitation energy and the oscillating strength, we tried Time-Dependent DFT calculation in Gaussian09 package[23]. Excited energies were calculated simultaneously up to 14th excitations.

The scaling factor "s" between the experimental energy to the TD-DFT calculated energy is important for the study of molecular excitation (absorption) bands. It was set to be s=1.00 (calculated wavelength /experimented wavelength) comparing experiments of fullerene cation $(C_{60})^+$ by Fulara et al.[13], also by Strelnikov et al.[25], and additionally comparing experiments of coronene cation $(C_{24}H_{12})^+$ by Ehrenfreund et al.[26] and Hardy et al.[27]. Detailed discussions will be done in later sections.

### 3. Neutral Fullerene ($C_{60}$)

Fundamental molecular properties of neutral fullerene ($C_{60}$) are important to study cationic fullerene and graphene molecules. HOMO (Highest Occupied Molecular Orbit), LUMO (Lowest Unoccupied Molecular Orbit) and Energy gap were obtained by DFT, which was compared with experiments [28), 29)] as shown in Fig.1. DFT employing PBE functional and 6-311G basis set gives HOMO to be -6.06eV, LUMO -4.30eV and energy gap 1.77eV, which were compared with experimental ones of -6.1eV, -4.3eV and 1.8eV respectively[28),29)]. Coincidence between them is excellent. Scaling factor of DFT was set to be s= 1.00. Accuracy of calculation was restricted by experimental accuracy. While in case of B3LYP functional and 6-311G basis set, calculated result was no good as compared in Fig. 1, no good HOMO and LUMO values and almost 1.6 times larger energy gap than experiments. We should apply PBE for both DFT and TD-DFT calculation.

By TD-DFT, molecular photoexcitation (absorption) energy in wavelength (Angstrom, with astronomical air incident correction) was obtained as shown in Table 1. From excitation number N1 to N14, wavelength range was optical from 7893A to 7046A. Unfortunately, all of the calculated oscillator strength was zero. We cannot compare with any experiments. Such zero value comes from very high symmetry of neutral fullerene (C60). As noted in Appendix 1, due to high molecular symmetry, sum of configuration interaction expansion coefficients becomes zero to give excitation strength in the Hartree-Fock determinant.

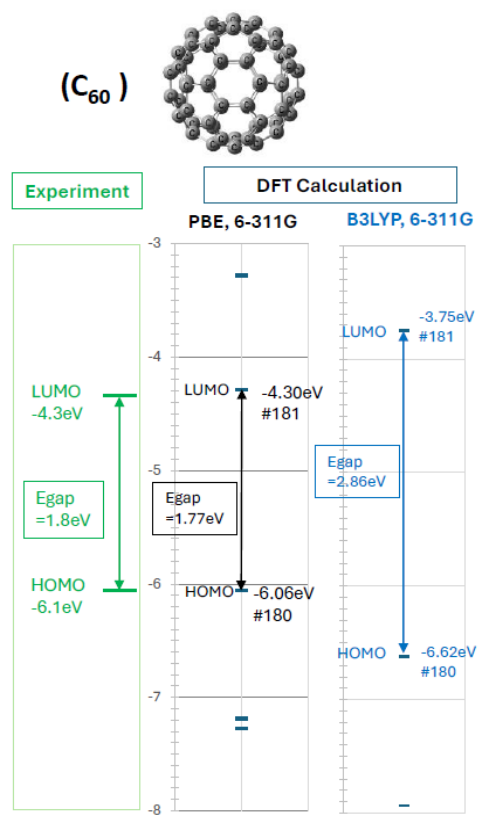

**Fig. 1** Calculated HOMO, LUMO and Energy gap of neutral fullerene ($C_{60}$) compared with experiments [28),29)]. DFT calculation employing PBE gives good coincidence rather than conventional B3LYP.

**Table 1** Calculated photoexcitation of neutral fullerene ($C_{60}$), excitation number N, wavelength in A, and the oscillator strength f. All oscillator strength was zero due to high symmetry of molecular configuration.

(C60), Charge=0, Sz=0/2

| Excitation Number | Wavelength in air (A) | Oscillator strength (10E-4) |
|---|---|---|
| N1 | 7892.92 | 0 |
| N2 | 7865.13 | 0 |
| N3 | 7762.76 | 0 |
| N4 | 7316.28 | 0 |
| N5 | 7315.18 | 0 |
| N6 | 7310.58 | 0 |
| N7 | 7309.68 | 0 |
| N8 | 7296.18 | 0 |
| N9 | 7257.39 | 0 |
| N10 | 7249.80 | 0 |
| N11 | 7248.19 | 0 |
| N12 | 7064.64 | 0 |
| N13 | 7062.74 | 0 |
| N14 | 7046.25 | 0 |

## 4. Fullerene Cation $(C_{60})^+$

### 4.1 HOMO and LUMO of fullerene cation $(C_{60})^+$

DFT calculated bands of fullerene cation $(C_{60})^+$ are illustrated in Fig. 2, which split to Alpha and Beta bands having doublet spin state of $Sz=1/2$. Comparing with neutral $(C_{60})$ in Fig. 1, one electron was pull out from Beta band resulting narrow energy gap of only 0.08eV. Such narrow gap may give near infrared and mid infrared photoabsorbance.

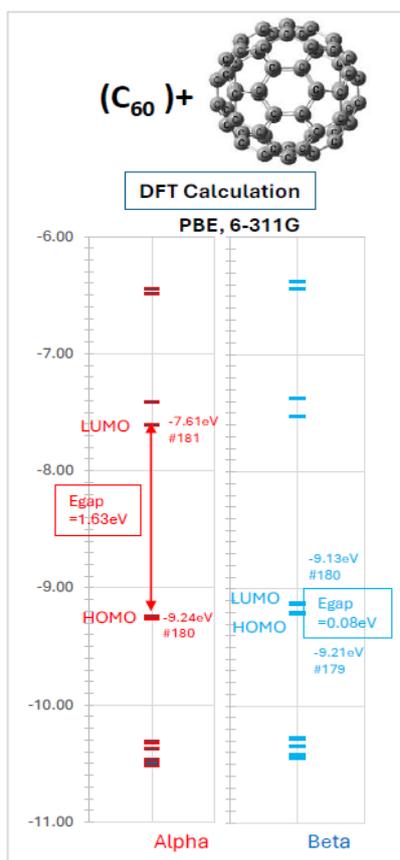

**Fig. 2** DFT calculated energy bands of Fullerene cation $(C_{60})^+$ splitting to Alpha and Beta bands. Energy gap of Beta band is small of 0.08eV.

### 4.2 Photoexcitation Experiments of $(C_{60})^+$

There are several experiments on photoexcitation of fullerene cation $(C_{60})^+$. Campbell et al. [16), 17)] (2015, 2016) confirmed absorbed bands including largest intensity two bands at 9577A and 9632A at low temperature gas phase environments as illustrated on (A) of Fig. 3 by light blue (only show wavelength position). Major five bands coincided very well with the astronomically observed DIBs. Preceding important experiment was done by Fulara et al. (1993) [13)] in Ne matrix shown in (B) by blue lines from 8300 to 9700A.

In the same paper, Fulara et al. did additional experiment in Ar matrix with doping $CCl_4$ to enrich cation ingredient $(C_{60})^+$ as noted on right columned of (C). Also, Strelnikov et al. [25)] (2015) did similar experiment in Ne matrix doping $1\%CO_2$ to enrich $(C60)^+$. Both doping experiments suggest longer wavelength band at 10620A and 10560A marked by red arrows respectively, which suggest unknown major bands of $(C_{60})^+$.

It was also interesting that shorter wavelength absorbed band at 8533A was recognized in Fulara experiment, also experimental band at 8500A by Strelnikov as marked by broken red dotted arrows in Fig. 3. Those may suggest additional unknown absorption bands of $(C_{60})^+$.

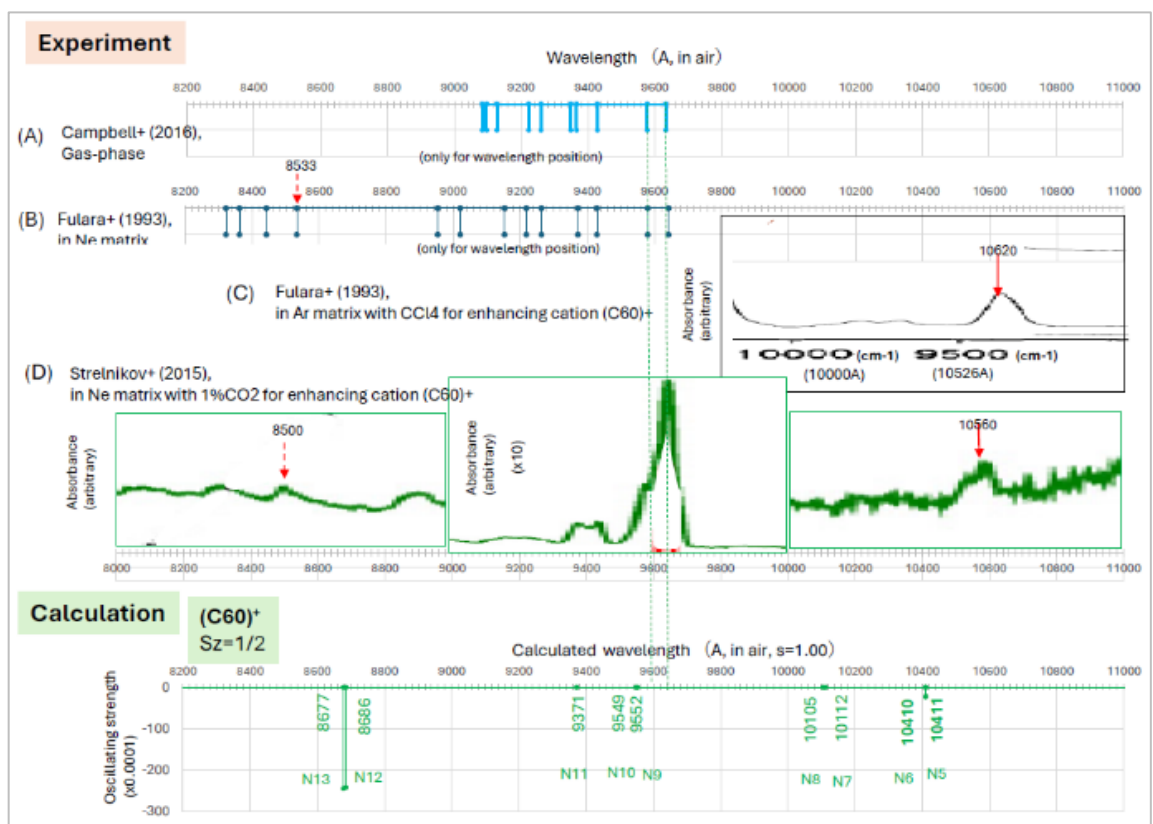

**Fig. 3** Photoexcitation of $(C_{60})^+$, (A) Gas-phase experiment by Campbell, (B) in Ne matrix by Fulara[13], (C) in Ar matrix with $CCl_4$ for enriching cation by Fulara[13], (D) in Ne matrix with 1%$CO_2$ enriching cation by Strelnnikov[25]. TD-DFT calculated bands were illustrated at the bottom by green lines.

**Table 2** Photoexcitation bands of $(C_{60})^+$. TD-DFT calculation on left column compared with experiments on right. Averaged scaling factor s=1.00.

| | TD-DFT Calculation PBE, 6-311G | | | Experiment (Wavelength, A in air) Scaling factor : s | | | | | |
|---|---|---|---|---|---|---|---|---|---|
| Excitation number | Wavelength (A, in air) | Oscillating strength | | Campbell (2016) gas-phase | s= Cal/Exp | Fulara (1993) Ne matrix | Fulara (1993) gas-phase with CCl4 | s= Cal/Exp | Strelnkov (2015) Ne matrix with CO2 | s= Cal/Exp |
| | Cal | | A | Cal/A | B | B | Cal/B | C | Cal/C |
| N1 | 148131.52 | 0 | | | | | | | |
| N2 | 146826.27 | 0 | | | | | | | |
| N3 | 140486.41 | 0 | | | | | | | |
| N4 | 139038.81 | 0 | | | | | | | |
| N5 | 10410.65 | 0.0022 | | | | | | | |
| N6 | 10409.95 | 0.0023 | | | | 10620 | 0.98 | 10560 | 0.99 |
| N7 | 10111.73 | 0 | | | | | | | |
| N8 | 10105.43 | 0 | | | | | | | |
| N9 | 9551.88 | 0 | 9632 | 0.99 | 9645 | | 0.99 | | |
| N10 | 9548.98 | 0 | 9578 | 1.00 | 9583 | | 1.00 | 9600 | 0.99 |
| N11 | 9371.53 | 0 | 9366 | 1.00 | 9374 | | 1.00 | | |
| N12 | 8685.61 | 0.0241 | | | | | | | |
| N13 | 8677.42 | 0.0244 | | | 8536 | | 1.02 | 8500 | 1.02 |
| N14 | 7994.10 | 0 | | | | | Average s=1.00 | | |

### 4.3 Photoexcitation bands of $(C_{60})^+$ by TD-DFT

Photoexcitation bands of $(C_{60})^+$ by TD-DFT calculation was shown in left column of Table 2. There are many coupled degenerated bands. Among 14 bands, there were four non-zero oscillating strength bands, which are coupled excitation numbers of N5 and N6, also coupled N12 and N13. Closest experimental band to calculated N5 (10411A) and N6 (10410A) was found at 10560A in an experimental figure by Strenikov[25] and at 10620A by Fulara[13]. Also, it is interesting that calculated N12 (8686A) and N13 (8677A) are close to experimental 8536A (Fulara) and around 8500A (Strelnikov). Under those comparison, we could set s= (calculated wavelength/experimental wavelength) as listed in Table 2 for every experiment. Averaged s value was s=1.00, which is the same with DFT. We should remind that the accuracy of calculated wavelength is 1%, limited by experimental accuracy. For example, in case of calculated 10000A wavelength, accuracy will be plus or minus 100A.

### 5, Astronomical Near Infrared DIB Bands

Recently, systematic near-infrared DIBs were opened by Hamano et al. [18] as shown in Fig. 4, and on Appendix 2. In Fig. 4, those observed DIBs are compared with TD-DFT calculated bands of $(C_{60})^+$.

### 5.1 Fullerene cation $(C_{60})^+$ related DIBs

In fig. 4, observed DIBs by Hamano et al.[18] were compared with calculated bands of fullerene cation (C60)+. We can review as follows,

(1) DIB9577 and DIB9632 are the DIBs specified to be $(C_{60})^+$ by Campbell et al.[16],[17]. In our calculation, coupled degenerated states of N10 (9549A) and N9 (9552A) are close to those DIBs. Unfortunately, those calculated oscillating strength are both zero, which should not correspond simply with experiment. The Jahn-Teller deformation[31] may give a good explanation as discussed in the next section.

(2) Calculated N6 (10410A) and N5 (10411A) suggest new DIBs related to $(C_{60})^+$, may correspond to DIB10439, DIB10542, and/or DIB10610.

(3) Calculated N12 (8686A) and N13 (8677A) suggest unknown DIBs belonging to $(C_{60})^+$.

### 5.2 Jan-Teller Deformation

In 2019, Lykhin et al. [30] proposed an attractive explanation on strong absorbance of DIB9577 and DIB9632 by the Jahn-Teller deformation [31], which causes remarkable band splitting on originally high symmetrical degenerated bands. They calculated energy difference between them to be 41cm-1, which is reasonably related to observed 60cm-1 as illustrated in Fig. 4.

Our simple TD-DFT calculation supposed no configuration change of ball like fullerene, resulting degenerated bands with zero oscillating strength due to very high symmetry. Recently in 2021, Nie et al. [32] confirmed strong correlation of DIB9577 and DIB9632 by comparing the astronomical equivalent width between them. They concluded that those two bands belongs to $(C_{60})^+$.

We compared observed DIBs by Hamano et al.[18] with experiment by Strelnikov et al. [25] as shown on Fig. 5. Green marked strong peak around 9600A looks split to two bands correspond to DIB9577 and DIB9632.

It is interesting that as illustrated in Fig. 5 there occurs similar coincidence at longer wavelength region. Experimental small peak around 10600A looks split to two bands just correspond to astronomical DIB10542 and DIB10610. Energy difference between them is 60cm-1, which is the same with the case of DIB 9577 and DIB9632. It is reasonable that once Jahn-Teller deformation occurs, there brings band splitting for all degenerated coupled bands.

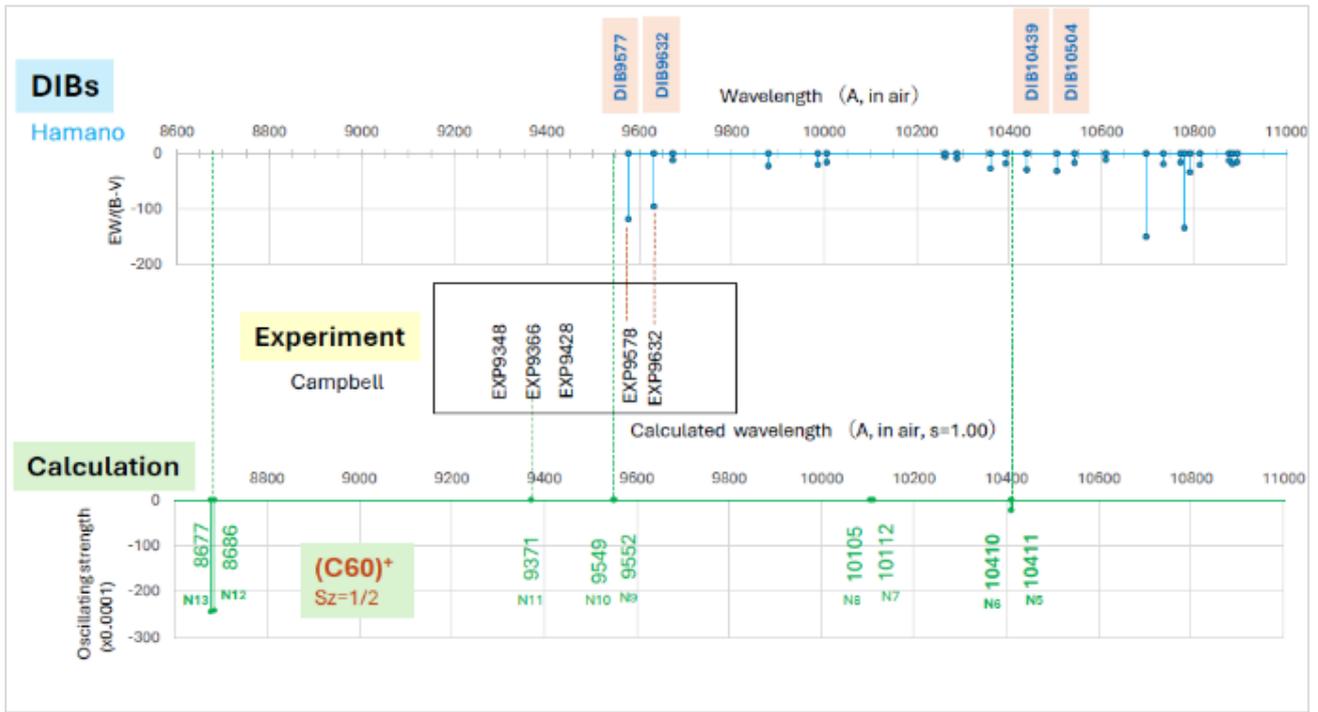

**Fig. 4** Near infrared observed DIBs by Hamano et al.[18] compared with calculation on cation $(C_{60})^+$.

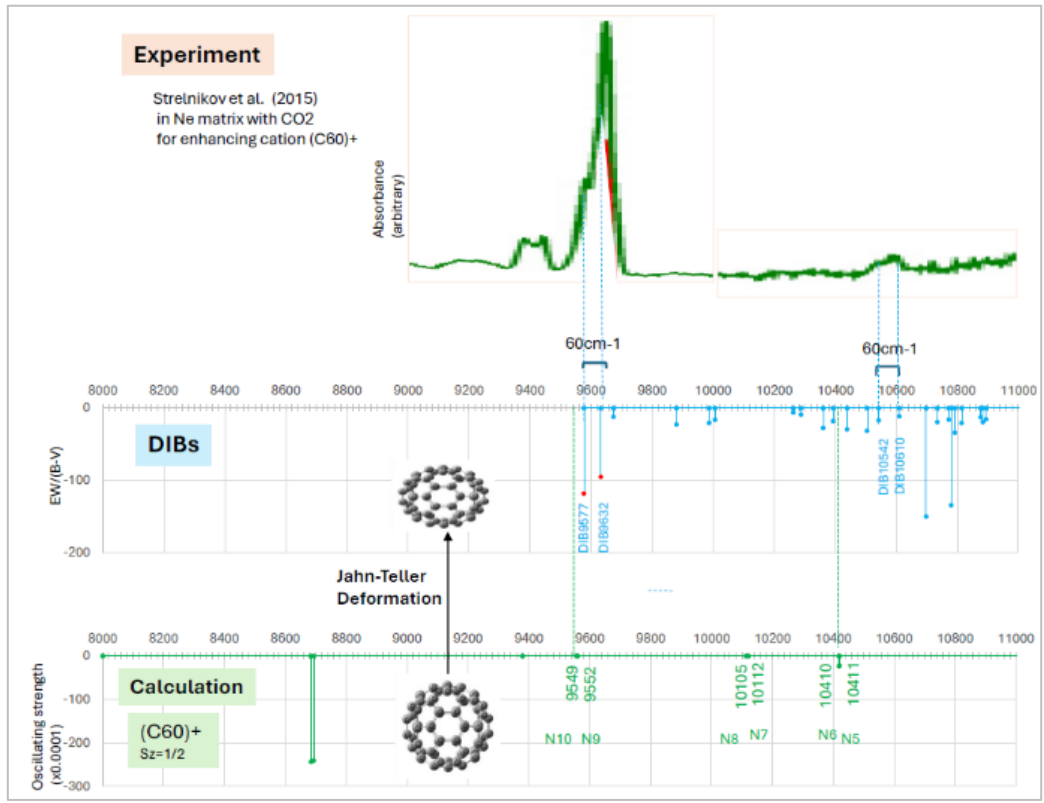

**Fig. 5,** Jahn-Teller deformation brings band splitting on both laboratory experiment and DIBs. Experimental strong peak around 9600A looks split to two bands correspond to DIB9577 and DIB9632. Also, a peak around 10600A looks split to two bands correspond to DIB10542 and DIB10610.

# 6, Graphene Molecules

## 6.1 Graphene molecules in interstellar space

In our previous papers [2),3)], it was emphasized that both fullerene and graphene contribute on the astronomical observed IRs. Those carbon molecules may float in interstellar space, also would be carrier candidates of DIBs.

As shown on left in Fig. 6, graphene molecule ($C_{54}$) with carbon hexagon networks would be attacked by high-speed proton or electron coming from the central star. One carbon will be kicked out resulting ($C_{53}$) having one pentagon site and 18 hexagon sites. Circled number ⑤ shows the pentagon site. Also, high energy photon irradiated from the central star may pull out one electron to transform neutral molecule to cation. Successive attack creates ($C_{52}$), ($C_{51}$) and so on. Side view of those molecules varies from plane to the cup like configuration. Finally, we can expect the creation of soccer ball like fullerene-($C_{60}$). Stable spin state of each molecule was calculated by DFT.

Band structure of neutral ($C_{53}$) was shown in Fig. 7 calculated by DFT. Stable spin state was *Sz=2/2*. Energy gap of Alpha band was 0.45eV, while Beta one 0.66eV. Every HOMO and LUMO orbitals are highly degenerated.

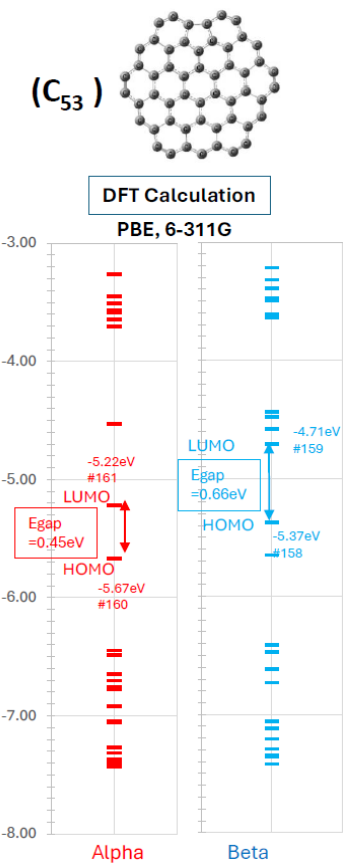

**Fig. 7** Band structure of neutral ($C_{53}$) with stable spin state of *Sz=2/2*.

**Fig. 6** Molecular candidates as carrier of DIBs. Molecular configuration, stable spin state was calculated for every neutral and cationic molecule by DFT.

## 6.2 Photoexcitation bands of Graphene molecule.

Photoexcited bands of graphene molecules were calculated by TD-DFT. Scaling factor was set to be s=1.00 based on $(C_{60})^+$ experiments [13), 25)], additionally checked by $(C_{24}H_{12})^+$ experiments [26), 27)] as noted in Appendix 3. Typical examples were neutral $(C_{54})$ and cation $(C_{54})^+$. It is interesting that nonzero oscillating strength bands arise for neutral $(C_{54})$ at N2 (10383A), and N6 (9634A). Also for cation $(C_{54})^+$, nonzero strength bands were at N2 (15504A), N3 (14934A), and N11 (10791A). It should be noted that graphene molecules arise many near-infrared bands. This is the first prediction. Graphene molecules may contribute on near-infrared DIBs.

In Fig. 8, we compared DIBs with calculated neutral graphene molecules of $(C_{54})$, $(C_{53})$, $(C_{52})$, and $(C_{51})$. We can see that there are many excited bands in near infrared region. Also, in Fig. 9, we compared with cationic graphene molecules. Again, we can see several bands in near infrared region.

**Table 3,** Calculated photoexcited bands for neutral $(C_{54})$ and cationic $(C_{54})^+$.

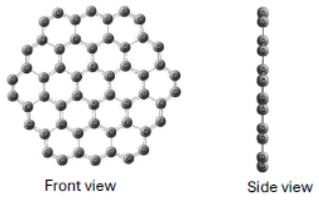

## 6.3 DIBs related to $(C_{54})$ and $(C_{54})^+$.

We like to compare DIBs with calculated photoexcited bands of $(C_{54})$ and $(C_{54})^+$. Accuracy of wavelength was estimated to be plus or minus 1%, that is, we should find specific DIB within an accuracy of plus or minus 100A.

For neutral $(C_{54})$ as illustrated in Fig. 8, nonzero oscillating strength band of N6 (9634A) is close to DIB9632, while degenerated states of N1 (10383A) and N2 (10383A) are close to DIB10394. Considering accuracy of calculation, one specified DIB among (DIB9577, DIB9632, DIB9673) may correspond to one of (DIB10361, DIB10394, DIB10439).

For cationic $(C_{54})^+$, calculated N11 (10753A) may correspond to one of many DIBs from 10700 to 10800A.

To identify specific DIB among those family, we need detailed astronomical study as like the equivalent width correlation [33)], also need detailed laboratory experiments.

## 6.4 DIBs related to $(C_{51})$ and $(C_{51})^+$.

Graphene $(C_{51})$ has three carbon pentagon sites and show many nonzero excited bands in near infrared region.

For neutral $(C_{51})$ as compared in Fig. 9, a band at N12 (9577A) looks related to one DIB among a group of (DIB9686, DIB9987, or DIB10006). Also, the band at N9 (10264A) may related to one of (DIB10262 or DIB10288).

For cationic $(C_{51})^+$, N5 (10234A) may correspond to one of (DIB10262 or DIB10288). Also, N4 (10767A) may correspond to one of many DIBs from 10690A to 10800A. Anyway, we need detailed astronomical study and laboratory experiments.

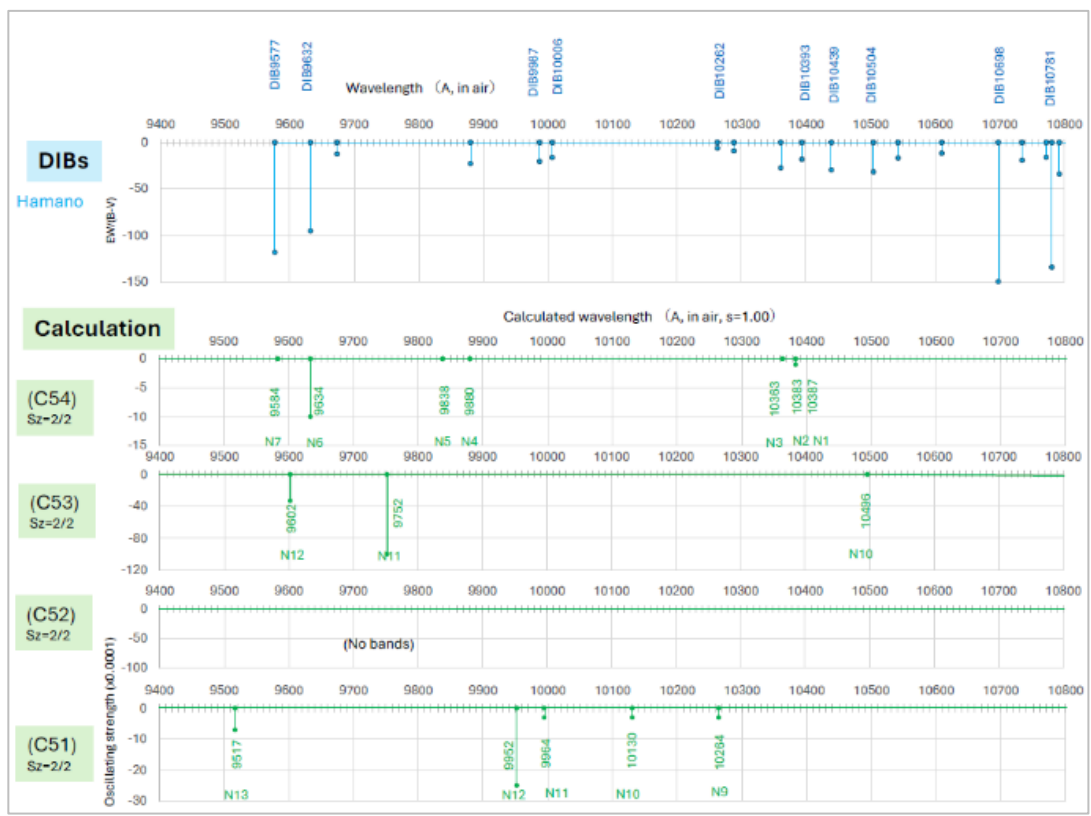

**Fig. 8** Near infrared absorbed bands of DIBs compared with calculated ones of charge neutral graphene molecules, $(C_{54})$, $(C_{53})$, $(C_{52})$ and $(C_{51})$.

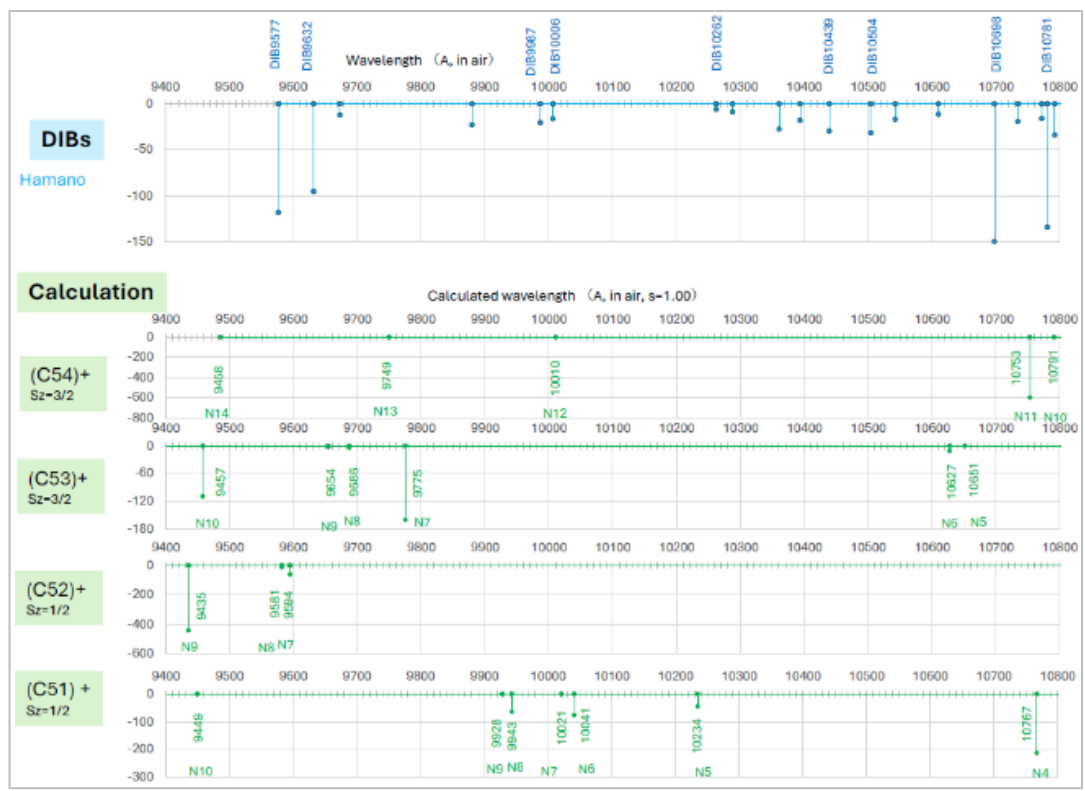

**Fig. 9** Near infrared DIBs compared with calculated cationic graphene molecules.

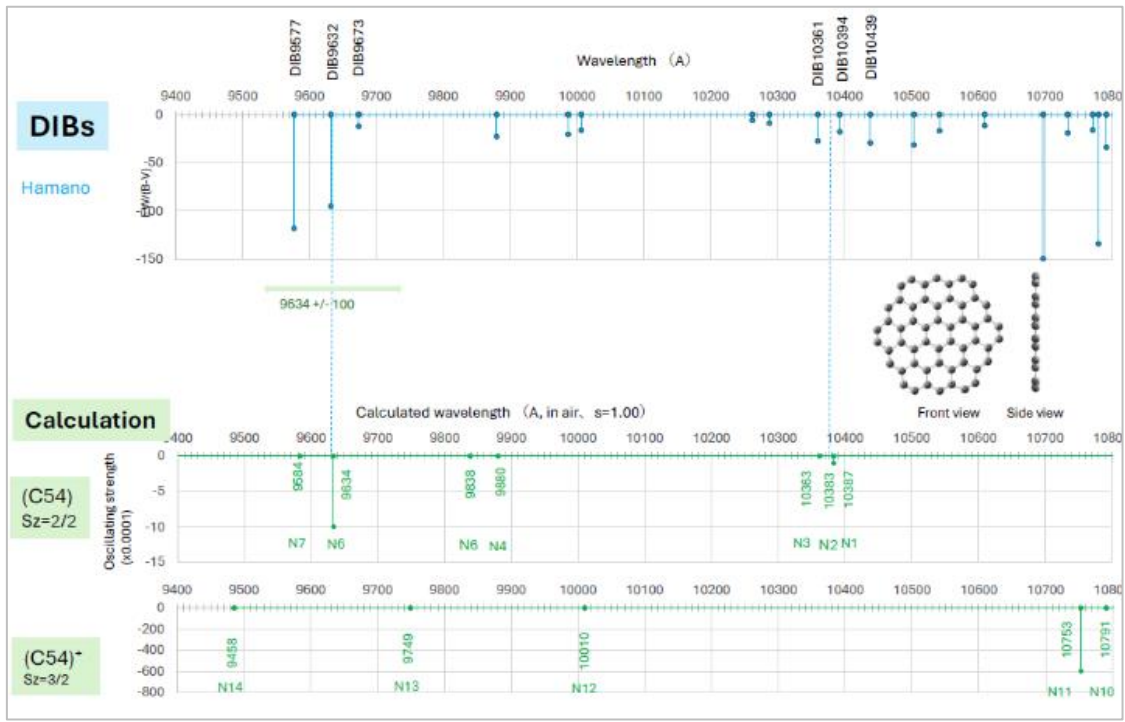

**Fig. 10**, Comparison of DIBs with calculated bands of $(C_{54})$ and $(C_{54})^+$. One specific DIB among (DIB9577, DIB9632, or DIB9673) may correspond to one of (DIB10361, DIB10394, or DIB10439).

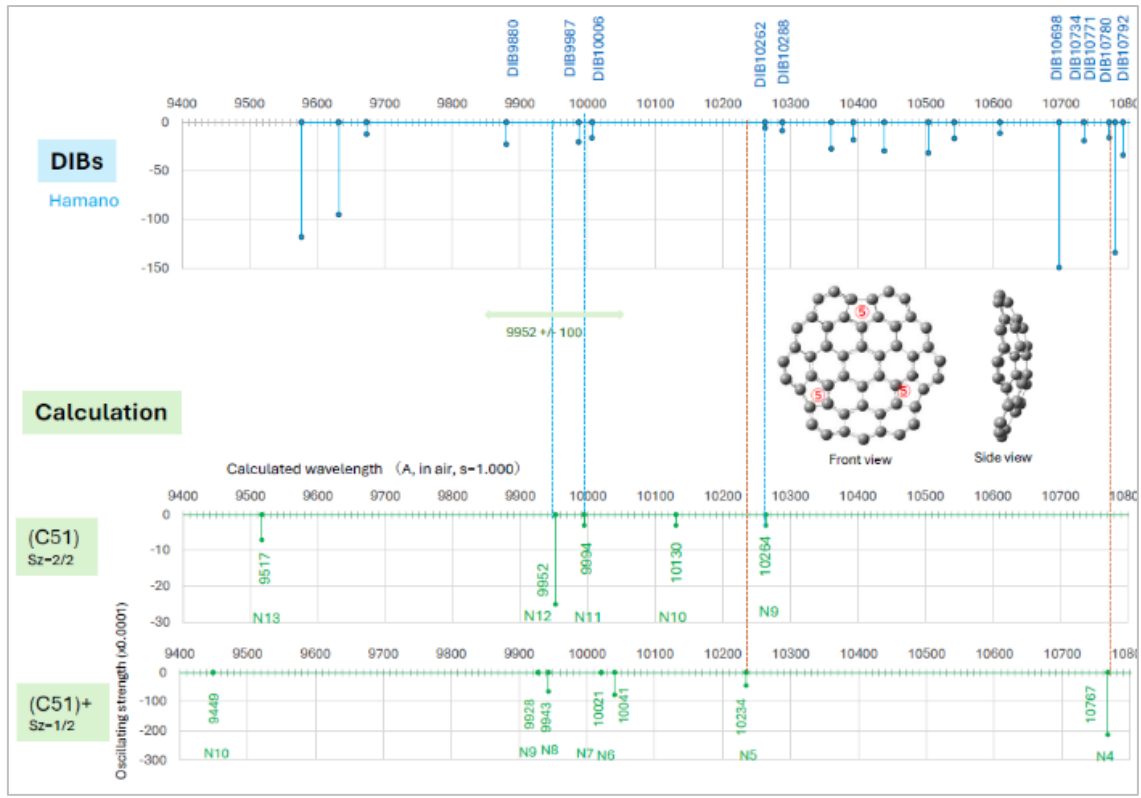

**Fig. 11**, Calculated band N12 (9577A) of $(C_{51})$ looks related to one of (DIB9686, DIB9987, or DIB10006), also N9 (10264A) to one of (DIB10262 or DIB10288). Similarly, $(C_{51})$+ suggests a family of DIBs.

# 7. Conclusion

Astronomical near-infrared DIBs were characterized by pure carbon fullerene and graphene molecules comparing with laboratory experiments and our TD-DFT calculation employing PBE functional 6-311G basis.

(1) Two large observed bands of fullerene cation $(C_{60})^+$ were coincide well with laboratory experimental bands by Campbell[16,17], Fulara[13] and Strelnikov[25]. In our TD-DFT calculation, those are reproduced by degenerated bands at 9549A and 9552A.
(2) Additional experiment of $(C_{60})+$ by Fulara[13] suggest a band around 10620A, also by Strelnikov a band at 10560A. Our TD-DFT calculation suggests degenerated bands of 10410A and 10411A.
(3) Shorter wavelength calculated bands of 8677A and 8686A may explain experimental 8533A by Fulara and 8500A by Strelnikov.
(4) Degenerated bands in TD-DFT could be split by the Jahn-Teller effect, which was theoretically explained by Lykhin et al.[30]. An experimental peak around 10520A by Strelnikov looks split to two DIBs of DIB10542 and DIB10610.
(5) Graphene molecules, as like $(C_{54})$, $(C_{53})$, $(C_{52})$, and $(C_{51})$, are carrier candidates for DIBs. TD-DFT calculation could show many near-infrared bands.
(6) Comparison of DIBs with calculated bands of $(C_{54})$ could suggest that one DIB among (DIB9577, DIB9632, DIB9673) may correspond to one of (DIB10361, DIB10394, DIB10439).
(7) Calculated bands of $(C_{51})$ suggest that one of (DIB9686, DIB9987, or DIB10006) may related to one of (DIB10262 or DIB10288).

This study pointed out that by combining three studies, that is, astronomical observation, laboratory experiment and quantum chemical analysis, we could suggest background carrier molecules.

# Acknowledgement

I would like to say great thanks to Prof. H. Nomura, Dr. S. Hamano, Dr. Y. Komatsu of National Astronomical Observatory Japan, and to Dr. M. Araki of Max Planck Institute, for sincere and deep discussions on DIB, continued efforts from January to April 2024.

---

Authors Profile  : Norio Ota, PhD. (太田憲雄)
Magnetic materials and physics,
Fellow, Honorable member of the Magnetics Society of Japan,
2010-2021: Senior Professor, Univ. of Tsukuba, Japan.
2003-2011: Executive Chief Engineer, Hitachi Maxell Ltd. Japan.
https://www.researchgate.net/profile/Norio-Ota-Or-Ohta

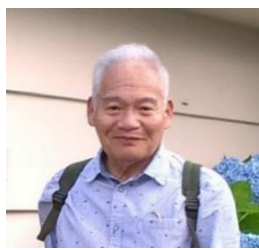

---

## Appendix 1.

## Appendix 2.

NIR Observed DIBs (Hamano)

| Wavelength(A) | EW/E(B-V) |
|---|---|
| 9577.00 | -118 |
| 9632.10 | -95 |
| 9673.40 | -12.2 |
| 9880.10 | -22.7 |
| 9987.00 | -20.4 |
| 10006.60 | -16.1 |
| 10262.50 | -6 |
| 10288.00 | -8.9 |
| 10360.70 | -27.4 |
| 10393.50 | -18 |
| 10439.00 | -29.4 |
| 10504.40 | -31.5 |
| 10542.60 | -16.8 |
| 10610.30 | -11.4 |
| 10697.60 | -149.5 |
| 10734.50 | -19.1 |
| 10771.90 | -16 |
| 10780.60 | -134 |
| 10792.30 | -33.9 |
| 10813.90 | -20.5 |
| 10876.90 | -12.7 |
| 10883.90 | -19.1 |
| 10893.90 | -15.5 |
| 11018.20 | -10.6 |
| 11691.60 | -17.2 |
| 11695.00 | -21 |
| 11698.50 | -29.5 |
| 11709.90 | -9 |
| 11720.80 | -38.7 |
| 11792.50 | -18.6 |
| 11797.50 | -119.8 |
| 11863.50 | -10.4 |
| 11929.30 | -14 |
| 12194.40 | -8.4 |
| 12200.70 | -17.7 |
| 12222.50 | -30.9 |
| 12230.00 | -14.3 |
| 12294.00 | -20.5 |
| 12313.50 | -7.2 |
| 12337.10 | -110.8 |
| 12519.00 | -21.1 |
| 12537.00 | -31.1 |
| 12594.90 | -10.7 |
| 12624.10 | -68.5 |
| 12650.00 | -19.7 |
| 12691.90 | -19.4 |
| 12798.80 | -35.4 |
| 12837.60 | -57.9 |
| 12861.50 | -38.3 |
| 12878.90 | -11.2 |
| 13021.00 | -14.6 |
| 13027.70 | -59.4 |
| 13021.00 | -14.6 |
| 13027.70 | -59.4 |
| 13050.50 | -11.7 |
| 13175.90 | -442.7 |

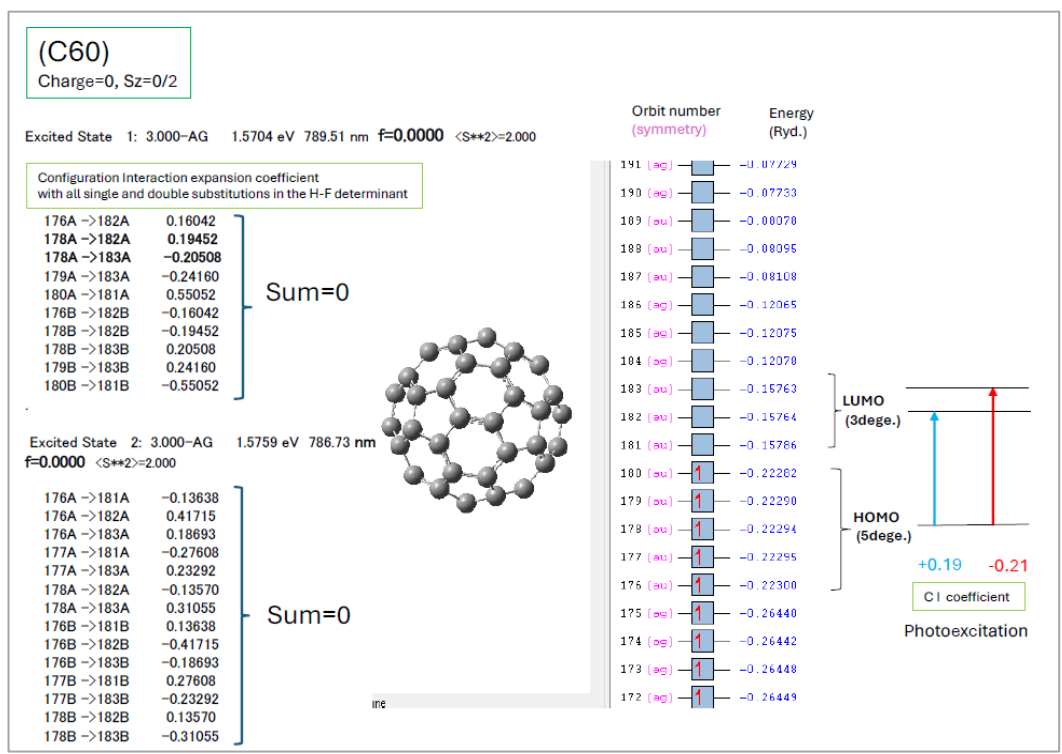

# Appendix 3

Calculated molecular orbital excitation wavelength    PBEPBE, 6-311Gbasis, s=1.000

### (C60), Charge=0, Sz=0/2

| Excitation Number | Wavelength in air (A) | Oscillator strength (10E-4) |
|---|---|---|
| N1 | 7892.92 | 0 |
| N2 | 7865.13 | 0 |
| N3 | 7762.76 | 0 |
| N4 | 7316.28 | 0 |
| N5 | 7315.18 | 0 |
| N6 | 7310.58 | 0 |
| N7 | 7309.68 | 0 |
| N8 | 7296.18 | 0 |
| N9 | 7257.39 | 0 |
| N10 | 7249.80 | 0 |
| N11 | 7248.19 | 0 |
| N12 | 7064.64 | 0 |
| N13 | 7062.74 | 0 |
| N14 | 7046.25 | 0 |

### (C60)+, Charge=+1, Sz=1/2

| Excitation Number | Wavelength in air (A) | Oscillator strength (10E-4) |
|---|---|---|
| N1 | 148131.52 | 0 |
| N2 | 146826.27 | 0 |
| N3 | 140486.41 | 0 |
| N4 | 139038.81 | 0 |
| N5 | 10410.65 | -22 |
| N6 | 10409.95 | -23 |
| N7 | 10111.73 | 0 |
| N8 | 10105.43 | 0 |
| N9 | 9551.88 | 0 |
| N10 | 9548.98 | 0 |
| N11 | 9371.53 | 0 |
| N12 | 8685.61 | -241 |
| N13 | 8677.42 | -244 |
| N14 | 7994.10 | 0 |

### (C54), Charge=0, Sz=2/2

| Excitation Number | Wavelength in air (A) | Oscillator strength (10E-4) |
|---|---|---|
| N1 | 11386.51 | 0 |
| N2 | 10383.49 | -1 |
| N3 | 10363.20 | 0 |
| N4 | 9880.33 | 0 |
| N5 | 9838.24 | 0 |
| N6 | 9633.80 | -10 |
| N7 | 9583.52 | 0 |
| N8 | 9287.50 | 0 |
| N9 | 9261.11 | 0 |
| N10 | 8981.78 | 0 |
| N11 | 8780.04 | 0 |
| N12 | 8721.86 | 0 |
| N13 | 6886.47 | -1 |
| N14 | 6434.20 | 0 |

### (C54)+, Charge=+1, Sz=3/2

| Excitation Number | Wavelength in air (A) | Oscillator strength (10E-4) |
|---|---|---|
| N1 | 113940.38 | 0 |
| N2 | 15504.25 | -70 |
| N3 | 14933.61 | -76 |
| N4 | 14545.41 | 0 |
| N5 | 13795.62 | 0 |
| N6 | 13652.66 | 0 |
| N7 | 13043.52 | 0 |
| N8 | 11881.14 | 0 |
| N9 | 11257.31 | 0 |
| N10 | 10791.04 | 0 |
| N11 | 10753.15 | -600 |
| N12 | 10010.25 | 0 |
| N13 | 9749.42 | 0 |
| N14 | 9485.10 | 0 |

### (C53), Charge=0, Sz=2/2

| Excitation Number | Wavelength in air (A) | Oscillator strength (10E-4) |
|---|---|---|
| N1 | 27561.15 | -19 |
| N2 | 19669.91 | 0 |
| N3 | 16559.46 | -1 |
| N4 | 14282.59 | -1 |
| N5 | 13619.77 | 0 |
| N6 | 13446.52 | 0 |
| N7 | 11861.95 | 0 |
| N8 | 11144.15 | -1 |
| N9 | 10855.83 | -2 |
| N10 | 10496.42 | 0 |
| N11 | 9752.33 | -100 |
| N12 | 9601.97 | -33 |
| N13 | 8725.90 | 0 |
| N14 | 8263.33 | -140 |

### (C53)+, Charge=+1, Sz=3/2

| Excitation Number | Wavelength in air (A) | Oscillator strength (10E-4) |
|---|---|---|
| N1 | 35162.67 | -1 |
| N2 | 23927.24 | -63 |
| N3 | 11988.52 | -41 |
| N4 | 11554.43 | 0 |
| N5 | 10651.48 | 0 |
| N6 | 10627.29 | -11 |
| N7 | 9775.12 | -161 |
| N8 | 9686.94 | -4 |
| N9 | 9654.15 | -3 |
| N10 | 9457.70 | -110 |
| N11 | 9386.22 | -1 |
| N12 | 9316.84 | -5 |
| N13 | 9300.35 | -74 |
| N14 | 9013.92 | 0 |

### (C52w), Charge=0, Sz=2/2

| Excitation Number | Wavelength in air (A) | Oscillator strength (10E-4) |
|---|---|---|
| N1 | 24831.69 | -1 |
| N2 | 17879.10 | -1 |
| N3 | 15112.76 | -1 |
| N4 | 13944.38 | -1 |
| N5 | 13184.09 | -3 |
| N6 | 12742.11 | -3 |
| N7 | 12614.54 | -4 |
| N8 | 9294.45 | -65 |
| N9 | 9214.07 | -18 |
| N10 | 8756.90 | -15 |
| N11 | 8399.50 | -91 |
| N12 | 8345.81 | -4 |
| N13 | 8219.35 | -38 |
| N14 | 7882.14 | -802 |

### (C52w)+, Charge=+1, Sz=1/2

| Excitation Number | Wavelength in air (A) | Oscillator strength (10E-4) |
|---|---|---|
| N1 | 23702.40 | -1 |
| N2 | 19253.72 | -95 |
| N3 | 14049.85 | -17 |
| N4 | 13511.20 | 0 |
| N5 | 13091.61 | -5 |
| N6 | 12771.20 | -2 |
| N7 | 9594.37 | -62 |
| N8 | 9581.37 | -14 |
| N9 | 9435.31 | -443 |
| N10 | 9214.37 | -177 |
| N11 | 8859.77 | -259 |
| N12 | 8436.99 | -56 |
| N13 | 8329.42 | -3 |
| N14 | 8305.52 | -2 |

### (C51t), Charge=0, Sz=2/2

| Excitation Number | Wavelength in air (A) | Oscillator strength (10E-4) |
|---|---|---|
| N1 | 103083.96 | -1 |
| N2 | 41157.92 | -11 |
| N3 | 29507.12 | -11 |
| N4 | 29347.26 | -12 |
| N5 | 16063.10 | 0 |
| N6 | 12115.68 | -238 |
| N7 | 11307.50 | -4 |
| N8 | 10928.50 | -196 |
| N9 | 10264.18 | -3 |
| N10 | 10130.82 | -3 |
| N11 | 9994.56 | -3 |
| N12 | 9952.27 | -25 |
| N13 | 9517.09 | -7 |
| N14 | 9352.73 | -14 |

### (C51t)+, Charge=+1, Sz=1/2

| Excitation Number | Wavelength in air (A) | Oscillator strength (10E-4) |
|---|---|---|
| N1 | 16500.38 | -4 |
| N2 | 16303.13 | -9 |
| N3 | 10802.94 | -172 |
| N4 | 10767.25 | -214 |
| N5 | 10234.70 | -45 |
| N6 | 10041.05 | -77 |
| N7 | 10021.15 | -1 |
| N8 | 9943.27 | -65 |
| N9 | 9928.07 | -1 |
| N10 | 9449.41 | -1 |
| N11 | 9193.58 | 0 |
| N12 | 8795.29 | -1 |
| N13 | 8702.71 | -18 |
| N14 | 8560.15 | -5 |

Appendix 4

## Coronene cation (C24H12)+
*Sz=1/2*

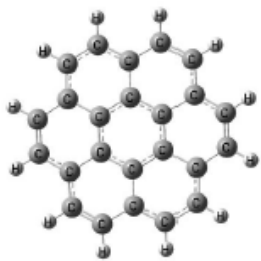

| Bare Calculation (A, in air) | | Experiment (A, in air) | | Experiment (A, in air) | |
|---|---|---|---|---|---|
| A | -f | E1 | A/E1 | E2 | A/E2 |
| Cal. | (x0.0001) | Ehrenfreunt (1992) *4) | s= Cal./Exp. | Hardy (2017) *5) | s= Cal./Exp. |
| 79573.90 | 0 | | | | |
| 9451.91 | -22 | 9465 | 0.999 | 9438 | 1.002 |
| 9267.56 | -87 | | | | |
| 7359.08 | -561 | | | | |
| 6872.71 | -206 | 6800 | 1.011 | | |
| 4736.80 | 0 | | | | |
| 4678.12 | -96 | | | | |
| 4612.63 | -95 | | | | |
| 4561.15 | 0 | 4590 | 0.994 | 4570 | 0.998 |
| 4394.19 | 0 | | | | |
| 4197.95 | -336 | | | | |
| 4145.66 | -107 | | | | |
| 3996.30 | 0 | | | | |
| 3915.32 | 0 | | | | |
| | | | $<s>=1.00$ | | $<s>=1.00$ |